\documentclass[useAMS,usenatbib]{mn2e}
\usepackage{graphicx,psfig,epsfig}

\def\lsim{\,\lower2truept\hbox{${<\atop\hbox{\raise4truept\hbox{$\sim$}}}$}\,}
\def\gsim{\,\lower2truept\hbox{${> \atop\hbox{\raise4truept\hbox{$\sim$}}}$}\,}

\title[Requirements for measuring CMB $B$-modes]{Foreground removal requirements for measuring large-scale CMB $B$-modes in light of  BICEP2}
\author[A.~Bonaldi, S.~Ricciardi and M.~L.~Brown]{A.~Bonaldi$^{1}$, S.~Ricciardi$^{2}$ and M.~L.~Brown$^{1}$\\
$^{1}$ Jodrell Bank Centre for Astrophysics, School of Physics \& Astronomy, University of Manchester, Oxford Road, Manchester M13 9PL, U.K.\\
$^{2}$INAF/IASF, Sezione di Bologna, Via Gobetti 101, I-40129 Bologna, Italy}
\begin{document}

\maketitle

\begin{abstract}
The most convincing confirmation that the $B$-mode polarization signal
detected at degree scales by BICEP2 is due to the Cosmic Microwave
Background (CMB) would be the measurement of its large-scale
counterpart. We assess the requirements for diffuse component
separation accuracy over large portions of the sky in order to measure the large-scale $B$-mode signal corresponding to a tensor to scalar ratio of $r=0.1$--0.2. 

We use the method proposed by Bonaldi \& Ricciardi (2011) to forecast the performances of different simulated experiments taking into account noise and foreground removal issues. We do not consider instrumental systematics, and we implicitly assume that they are not the dominant source of error. If this is the case, the confirmation of an $r=0.1$--0.2 signal is achievable by {\it Planck} 
even for conservative assumptions regarding the accuracy of
foreground cleaning. Our forecasts suggest that the combination of
this experiment with BICEP2 will lead to an improvement of 25--45\,\%
in the constraint on $r$. 

A next-generation CMB polarization
satellite, represented in this work by the COrE experiment, can reduce
dramatically (by almost another order of magnitude) the uncertainty on $r$. In this case, however, the accuracy of foreground removal becomes critical to fully benefit from the increase in sensitivity. 
\end{abstract}

\section{Introduction}
Measurements of the $B$-mode (curl component) polarization of the CMB provide a unique opportunity to detect the imprint of the primordial gravitational waves predicted by the inflationary paradigm. The amplitude of these tensor perturbations measures the energy scale of
inflation and its potential. Such measurements therefore can be used
to place powerful constraint on a broad class of inflationary models. Moreover, the confirmation of inflation and the determination of the inflationary potential would provide a direct observational link with the physics of the early universe.

The recent detection of a non-zero $B$-mode power spectrum at multipoles
around $\ell \sim 100$ by BICEP2 \citep{2014arXiv1403.3985B} have
profound implications for current and future experiments aimed at
measuring CMB polarization. If the signal detected by BICEP2 is confirmed as primordial, the implied relatively large value of the tensor-to-scalar ratio ($r \sim 0.2$) re-evaluates the requirements for both the instrument (sensitivity, systematics) and the data analysis (in particular foreground removal).

Given that foreground emission and instrumental systematics can
generate $B$-modes of significant power over broad multipole ranges,
it is possible that the origin of at least part of the signal measured
at degree scales by BICEP2 is not cosmological. The most convincing
confirmation would be a measurement of the reionization bump at
multipoles $\ell \sim 2$--10, and in general of the $B$-mode signal
over a wider range of angular scales. In fact, measuring the power spectrum in two different multipole regimes would probe its shape, which is expected to be very different from that due to foregrounds and/or instrumental systematics. 
Going to larger scales, and thus to the reionization bump, is probably the easier option, because the $B$-mode signal due to lensing dominates over the primordial one at $\ell > 100$. Probing the $B$-mode power spectrum over a wide multipole range would also improve considerably the measurement of the tensor-to-scalar ratio, $r$, and of the optical depth to reionization, $\tau$.

Measuring the $B$-mode power spectrum at the largest scales requires a
large sky coverage, ideally a full-sky satellite experiment. Foreground contamination must be dealt with while at the
same time retaining as much sky area as possible, thus foreground
avoidance (analyzing a small sky area where foreground emission is
particularly low, as done by BICEP2) is not a viable option. The
strategy in this case is to map the total polarization signal at
several frequencies and to exploit the different dependencies on frequency of the emission components to separate them. This data analysis step is called component separation. 

Several component separation approaches have been developed for CMB
$B$-mode detection at large and intermediate scales
\citep{2009A&A...503..691B,2009MNRAS.397.1355E,2010MNRAS.406.1644R,2010MNRAS.408.2319S,2011ApJ...737...78K,2011MNRAS.418.1498A,2013MNRAS.435...18B}. In
most cases, the process can be thought of as forming a suitable
combination of the data at different frequencies. Such a combination
is designed to minimize the foreground contribution while also
reducing the instrumental noise. However, the final noise level will
always be higher than what could be achieved in the absence of foregrounds, for example by averaging the data at different frequencies with inverse noise variance weights \citep{2011MNRAS.414..615B}. Thus, the optimization of an experiment targeting $B$-mode measurements on large scales needs to consider, together with the signal-to-noise ratio, also the issues related to foreground contamination and component separation \citep{2011MNRAS.414..615B,2011JCAP...08..001F,2012MNRAS.424.1914A,2012PhRvD..85h3006E}. 

In this work we use the forecasting tool developed in \cite{2011MNRAS.414..615B} to assess the component separation requirements to measure the large-scale $B$-mode signal for values of $r$ between 0.1 and 0.2, consistent with the detection by BICEP2. This tool estimates the uncertainties (noise, foreground residuals and cosmic variance) on the $B$-mode power spectrum taking into account the instrumental specifications (sensitivity, number and frequency of the channels, and sky coverage) and a component separation step (the CMB is obtained with a suitable linear combination of the frequency maps). We perform a Fisher matrix analysis to propagate the predicted uncertainties from the power spectrum to the cosmological parameters, in particular $r$ and $\tau$. 
We apply our method to the specifications of the {\it Planck}
satellite \citep{2013arXiv1303.5062P} and the Cosmic Origins Explorer
\citep[COrE,][]{2011arXiv1102.2181T}. We also consider a balloon-borne
experiment targeting the large-scale $B$-mode signal, for which we
take the specifications of the Large-Scale Polarization Explorer
\citep[LSPE,][]{2012arXiv1208.0281T}, as an example case.    

\section{Statement of the problem}
\label{sec:due}
\subsection{Data model}
In addition to the CMB, the microwave sky contains several foreground components, both diffuse and compact. For our analysis, which is focused 
on large and intermediate scales, we consider only diffuse foregrounds. The main diffuse polarized foregrounds are Galactic synchrotron 
and thermal dust (the free-free emission is unpolarized and the anomalous dust emission is also expected to be essentially unpolarized).

The synchrotron component dominates at the lowest frequencies. Its spectral behavior in antenna temperature can be modeled as a power law:
\begin{equation}\label{scaling_synchro}
T_{\rm A,synch}(\nu)\propto \nu^{-\beta_{\rm s}} \label{syn}\ ,
\end{equation}
where the synchrotron spectral index $\beta_{\rm s}$ can vary in the sky in the range $2.5<\beta_{\rm s}<3.5$. The spectral behaviour of thermal dust emission, which dominates at high frequencies, follows approximately a grey-body law:
\begin{equation}\label{scaling_dust}
T_{\rm A,dust} (\nu)\propto \frac{\nu^{\beta_{\rm d}+1}}{\exp
(h\nu/kT_{\rm dust})-1}. \label{dust}
\end{equation}
Both $\beta_{\rm d}$ and $T_{\rm d}$ are spatially varying around $\beta_{\rm d} \sim 1.7$ and $T_{\rm d} \sim 18\,$K. The polarized CMB signal has a blackbody spectrum:
\begin{equation}\label{scaling_CMB}
T_{\rm A,CMB}(\nu)\propto\frac{(h\nu/kT_{\rm CMB})^2\exp (h\nu/kT_{\rm CMB})}
{(\exp (h\nu/kT_{\rm CMB})-1)^2} \
\end{equation}
with $T_{\rm CMB}\simeq 2.73$.

For component separation purposes, it is convenient to model the data as a linear mixture of the components. For each direction in the sky we write  
\begin{equation}\label{datamodel}
\bmath{x}=\bmath{\sf{H}}\bmath{s}+\bmath{n}.
\end{equation}
The vectors $\bmath{x}$ and $\bmath{n}$ have dimension equal to the number of detectors, $N_{\rm d}$, and contain the data and instrumental noise, respectively;  
$\bmath{s}$ is a vector containing the sources (CMB and foregrounds) and has dimension equal to the number of components, $N_{\rm c}$; 
$\bmath{\sf{H}}$ is the $N_{\rm d} \times N_{\rm c}$ mixing matrix, containing the frequency scaling of the components. 
The spatial variability of the synchrotron and dust spectral indices implies that the mixing matrix $\bmath{\sf{H}}$ is in general different for different sky pixels.

In order to be able to write the data model as in Eq.~(\ref{datamodel}) we made some simplifying assumptions. The most important of them 
is that the instrumental resolution does not depend on frequency. This is in general not true, and requires a pre-processing step in which the resolution is equalised by suitably smoothing the data. In our case, because we focus on large scales, such a loss of resolution is not particularly problematic. 

If the linear mixture data model holds, the components can be reconstructed as
\begin{equation}\label{recon}
\bmath{\hat s}=\bmath{\sf{W}}\bmath{x},
\end{equation}
where  $\bmath{\hat s}$ is an estimate of the components $\bmath{s}$
and $\bmath{\sf{W}}$ is a $N_{\rm c} \times N_{\rm d}$ matrix called
the reconstruction matrix. We choose to rely on a linear estimator
because it will allow us to easily include the component separation
process in the forecasting of $B$-mode power spectrum constraints, as
we will see in the next section. In addition this approach is well
suited for use with Monte Carlo simulations, needed to accurately control error sources.

We adopt the so-called Generalized Least Square solution (GLS):
\begin{equation}
\label{gls}
\bmath{\sf{W}}=[\bmath{\sf{\hat H}}^T \bmath{\sf{N}}^{-1}\bmath{\sf{\hat H}}]^{-1}\bmath{\sf{\hat H}}^T\bmath{\sf{N}}^{-1}.
\end{equation}
This requires the noise covariance $\bmath{\sf{N}}$ of the channel maps and an estimate $\bmath{\sf{\hat H}}$ of the mixing matrix $\bmath{\sf{H}}$ 
that can be obtained exploiting any of the dedicated component
separation methods discussed in the literature \citep[see,
  e.g.,][]{2006MNRAS.373..271B,2006ApJ...641..665E,2009MNRAS.392..216S,2010MNRAS.406.1644R,2011MNRAS.418.1498A}. We
stress that this choice is  not completely general, as other
reconstruction matrices could be used. However, rather than on the
form of the reconstruction matrix, the main dependence is on the
mismatch between the true and the estimated mixing matrix. The fact
that our reconstruction matrix explicitly contains the estimated
mixing matrix also allows us to easily include in our forecasts the effect of errors in the mixing matrix estimation.
\begin{table*}
\centering
\begin{minipage}{140mm}
\caption{Instrumental characteristics considered in the present study
  for the {\it Planck} and COrE experiments. The RMS per pixel
  represents the polarization sensitivity and is quoted for Healpix
  resolution $N_{\rm side}=512$ pixels (pixel size $\sim 7\,$arcmin). The integration time is 4 years for COrE, 51 months for {\it Planck} LFI, 29 months for {\it Planck} HFI and two weeks for LSPE.} \label{tab:instr}
\begin{tabular}{llllllllllllllllll}
\hline
{{\it Planck} LFI \footnote{\cite{2013arXiv1303.5063P}} and HFI \footnote{\cite{2013arXiv1303.5067P}} specifications}\\
\hline
$\nu\,$(GHz)&44&70&100&143&&&217\\
FWHM (arcmin)&24&14&9.5&7.1&&&5.0\\
RMS $\Delta T$($\mu$K RJ)& 34 &38 & 11 &4 &&&2.9\\
\hline
COrE specifications \footnote{\cite{2011arXiv1102.2181T}} \\
\hline
$\nu\,$(GHz)&45&75&105&135&165&195&225\\
FWHM (arcmin)&23&14&10&7.8&6.4&5.4&4.7\\
RMS $\Delta T$($\mu$K RJ)&1.2&0.6&0.5&0.4&0.3&0.3&0.2\\
\hline
LSPE specifications \footnote{\cite{2012arXiv1208.0281T}} \\
\hline
$\nu\,$(GHz)&43&90&95&145&&&225\\
FWHM (arcmin)&60&30&110&89&&&74\\
RMS $\Delta T$($\mu$K RJ)&23&87&1.6&1.7&&&1.8\\
\hline
\end{tabular}
\end{minipage}
\end{table*}

\section{Forecast method}
\subsection {Errors on the $B$-mode power spectrum}\label{sec:errbar}
In this section we briefly summarise the forecasting method presented
in \cite{2011MNRAS.414..615B}. We refer the reader to this paper for 
further details and a complete derivation. In addition to considering the sky coverage of each instrument, we adopt for the analysis a Galactic mask to exclude those regions which are most contaminated by foreground emission. After masking we are left with a sky fraction $f_{\rm_sky}$. 

We assume the analysis will recover the CMB component from the
multi-frequency data through the GLS linear mixture estimator [eqns.~(\ref{recon}) and (\ref{gls})]. 
We use uniform weights across the considered sky area, which allows us
to work directly at the power spectrum level. This is not a realistic
assumption, because the spectral properties of the foreground
components are known to vary with  position on the sky. However, we
can still simulate the correct level of foreground contamination provided that the constant weights that we use are a good representation of the true ones over most of the sky. 

The other crucial parameter is the difference between the true mixing
matrix $\bmath{\sf H}$ and the estimated mixing matrix $\bmath{\hat
  {\sf H}}$. This difference needs to represent the estimation error
that we would have in an analysis of real data, i.e. when the spectral properties of the foregrounds are spatially varying. It can also be increased to incorporate other effects, such as an incorrect modelling of the foreground properties (e.g. steepening of the synchrotron spectral index) or the presence of additional polarised foregrounds.
In our case we parametrise the mixing matrix error in terms of
uncertainties on the synchrotron and dust spectral indices, $\Delta
\beta_{\rm s}$ and $\Delta \beta_{\rm d}$. As detailed in
Sect.~\ref{sec:deltabeta}, we consider two quite different error
regimes, which reflect our uncertainty on the current models and on
the progress we expect to make in the future.

We estimate the power spectrum in multipole bands  $\hat \ell$, according to some binning scheme. In the following we adopt the notation $\bmath{C}^{\rm XX}_{\hat \ell}$ where XX can be either the $EE$ or $BB$ CMB polarization spectrum.  We model the error $\Delta \bmath{C}^{\rm XX}_{\hat \ell}$ on the power spectrum $\bmath{C}^{\rm XX}_{\hat \ell}$ as the sum of three contributions: noise, residual foreground contamination, and cosmic variance
\begin{equation}
\label{total}
\Delta \bmath{C}^{\rm XX}_{\hat \ell}=\Delta \bmath{C}^{\rm XX}_{\hat \ell,{\rm noise}}+\Delta \bmath{C}^{\rm XX}_{\hat \ell,{\rm foreg}} + \Delta \bmath{C}^{\rm XX}_{\hat \ell,{\rm CV}}.
\end{equation}
By adding the three error components in this way we implicitly assume that the errors are Gaussian and uncorrelated. This might not be true
especially at the lowest multipoles, where the error distribution for
a given spectral bin can be non-Gaussian and highly
asymmetric. Moreover, we do not consider any correlation between the
errors on different bins or between the $EE$ and $BB$
bandpowers. Finally, we note that we are not considering here any
contribution due to instrumental systematics. Although systematics can
be very important for $B$-mode detection,  they are typically
instrument-specific, and cannot be predicted easily without modelling
the instrument in detail. For all the reasons listed above, our forecasted errors should be considered as an approximate, and possibly optimistic, assessment of the true uncertainties. 

Concerning the noise error, we do not know the actual noise realization but we can estimate its statistical properties trough a Monte Carlo analysis. The noise component of the error on the CMB power spectrum, $\Delta  \bmath{C}^{\rm XX}_{\hat \ell,{\rm noise}}$, is due to the sampling variance of the noise bias,
\begin{equation}
\label{deltanoise}
\Delta  \bmath{C}^{\rm XX}_{\hat \ell,{\rm noise}}=\sqrt{\frac{2/(2\hat \ell+1)}{f_{\rm sky} \, \bmath{\rm nbin}(\hat \ell)}}\,\bmath{N}_{\rm CMB},
\end{equation}
where $\bmath{N}_{\rm CMB}$ is the noise bias on the CMB power
spectrum and $\bmath{{\rm nbin}}(\hat \ell)$ contains the number of
multipoles within each of the spectral bins $\hat \ell$. If the noise is white and Gaussian, for each frequency $\nu$ we have
\begin{equation}\label{clnoise}
\bmath{N}_{\nu}=\frac{4 \pi f_{\rm sky}}{N_{\rm pix}} \sigma^2_{\nu}\bmath{B}_{\nu}^2
\end{equation}
where $N\rm {pix}$ is the number of pixels in the considered portion
of the sky $f_{\rm sky}$, $\sigma^2_{\nu}$ is the noise variance per pixel at frequency $\nu$ and
 $\bmath{B}_{\nu}^2(\ell)$ is the beam function applied to each channel map to obtain a common resolution. 

Given the linearity of the CMB recovery process [eq.~(\ref{recon})] and the assumption of a spatially-invariant reconstruction matrix, the noise bias 
is obtained by combining the channel noise spectra $\bmath{N}_{\nu}$ with the matrix $\bmath{\sf{W}}^2$:
\begin{equation}\label{clnoise}
\bmath{N}_{\rm CMB}=\sum_\nu w^2_{\nu,{\rm CMB}}\, \bmath{N}_{\nu},
\end{equation}
where $w^2_{\nu,{\rm CMB}}$ are the elements of the matrix $\bmath{\sf{W}}^2$ pertaining to the CMB component. From eqns. (\ref{deltanoise}) and (\ref{clnoise}) we see that the error due to noise depends on the reconstruction matrix $\bmath{\sf{W}}$ and, therefore, on the estimated mixing matrix $\bmath{\sf{\hat H}}$. It is possible that the cleanest channels in terms of instrumental noise are not the best in terms of foreground contamination. Thus, the optimal reconstruction matrix does not necessarily minimise noise. In particular, depending on the relative sensitivity of the frequency channels, it is possible that a very accurate mixing matrix corresponds to a noise level that is higher than that achieved with a less accurate mixing matrix.

The map of residuals, $\bmath{s}-\bmath{\hat s}$, for a linear mixture source reconstruction can be estimated as:
\begin{equation}
\label{deltacomp}
\bmath{s}-\bmath{\hat s}=(\bmath{\sf{W}}\bmath{\sf{H}}-\bmath{\sf{I}})\,\bmath{\tilde s},
\end{equation}
where $\bmath{\sf{I}}$ is the identity matrix and  $\bmath{\tilde s}$
is a set of simulated components
\citep{2010MNRAS.408.2319S,2011MNRAS.414..615B,2012PhRvD..85h3006E}. The
error due to the imperfect foreground subtraction, $\Delta C^{\rm
  XX}_{\hat \ell,{\rm foreg}}$, is the binned power spectrum of the
residuals computed outside the adopted Galactic mask. This error
essentially depends on the mismatch between the true mixing matrix
$\bmath{\sf{H}}$ and the estimated mixing matrix $\bmath{\sf{\hat
    H}}$, which is used to compute the reconstruction matrix
$\bmath{\sf{W}}$.

It is clear that $\Delta \bmath{C}^{\rm XX}_{\hat \ell,{\rm foreg}}$ is model-dependent, since it relies on simulations of the data $\bmath{\tilde s}$, which are hampered by our poor knowledge of polarized foregrounds. The situation will substantially improve in the very near future as new polarization data, above all those from the {\it Planck} mission, will become available. 

Finally, the cosmic variance term is given by
\begin{equation}
\label{deltacv}
\Delta  \bmath{C}^{\rm XX}_{\hat \ell,{\rm CV}} =\sqrt{\frac{2/(2\hat \ell+1)}{f_{\rm sky} \, \bmath{\rm nbin}(\hat \ell)}}\,\bmath{C}^{\rm XX}_{\hat \ell},
\end{equation}
and represents the error due to the fact we only measure one
particular CMB realization. It depends on the area of the sky
considered. We note that this formula is a good approximation only if
the fraction of the sky is large. Our forecasted errors for the LSPE
experiment, which we take to be representative of a large-scale
balloon-borne experiment (see Sect.~\ref{sec:inst_spec}), are therefore
likely to be underestimates. 

\subsection{Errors on the cosmological parameters} \label{sec:fisher}
We propagate the total error on the power spectrum  $\Delta \bmath{C}^{\rm XX}_{\hat \ell}$ to the cosmological parameters with a Fisher matrix approach.  
Given a set of parameters $p=\{p_1, p_2, ..p_N\}$ which depend on a set of observables $d=\{d_1, d_2, ..d_m\}$, the Fisher information matrix is a $N \times N$ symmetric matrix whose elements are given by
\begin{equation}
F_{ij}=\sum _{l=1,m}\frac{1}{\sigma^2_l}\frac{\partial d_l}{\partial p_i}\frac{\partial d_l}{\partial p_j},
\end{equation}
where $\sigma^2_l$ is the variance of the error on the datapoint $d_l$, and $\partial d_l/\partial p_i$ is the partial derivative of $d_l$ with respect to $p_i$. 
The inverse of the Fisher matrix, $F^{-1}$ is the covariance matrix for the parameters $p_i$; in particular, it contains the variance on the parameters on the diagonal $F^{-1}_{ii}=\sigma^2_{ii}$. 

In our case, the parameters $p_i$ are the cosmological parameters, and the variances $\sigma^2_l$ are the forecasted errors on the CMB $EE$ and $BB$ power spectra for a set of relevant bins $\hat \ell=1,m$.  The partial derivatives $\partial d_l/\partial p_j$ are evaluated numerically by computing power spectra from theory for a fiducial cosmological model. We then vary by a small amount one parameter at a time and  evaluate the corresponding change in the data $d_l$. 

\section{Details of the simulation}\label{sec:quattro}
\begin{figure*}
\begin{center}
\includegraphics[width=4.5cm,angle=90,keepaspectratio]{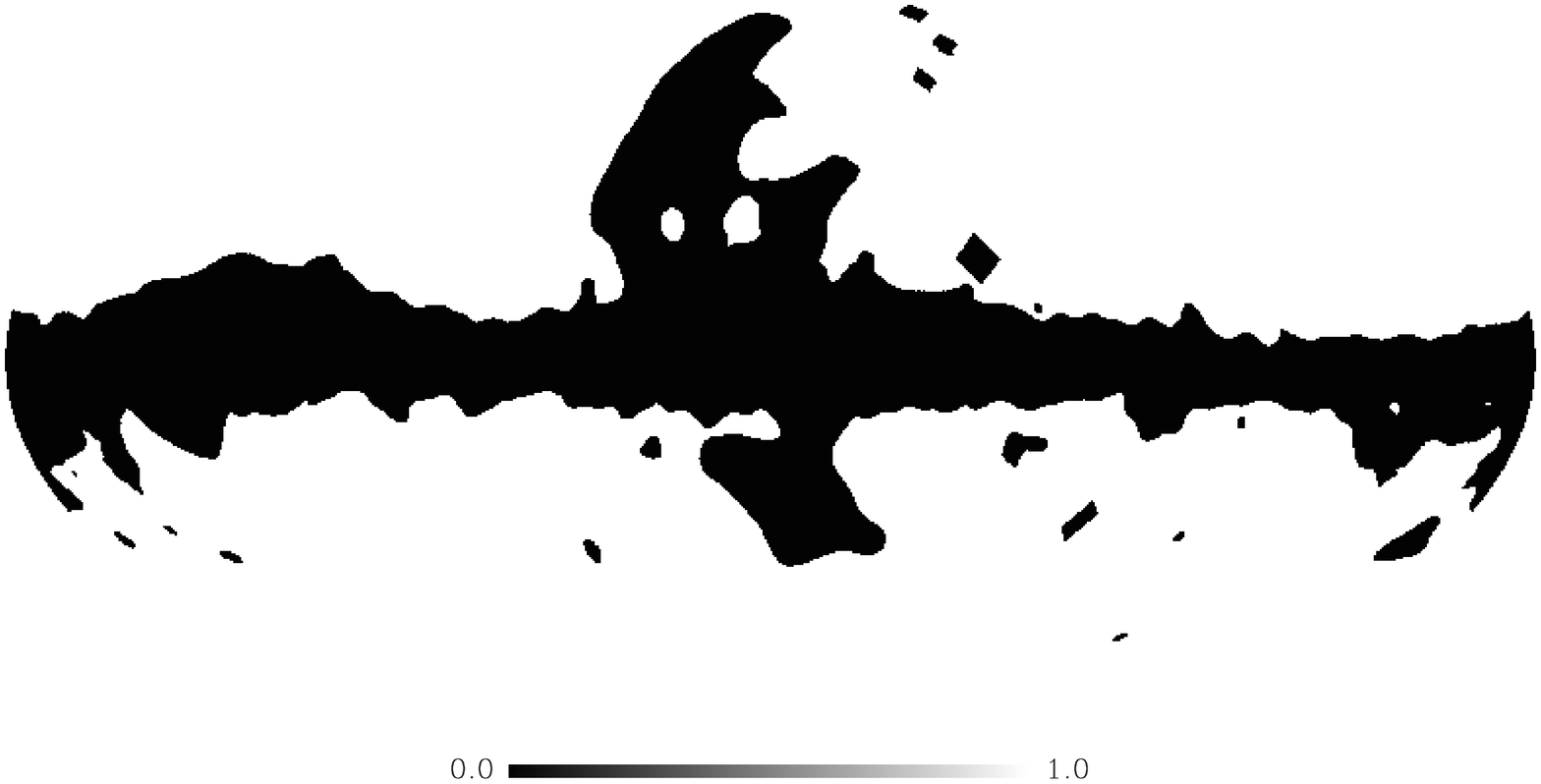}
\includegraphics[width=4.5cm,angle=90,keepaspectratio]{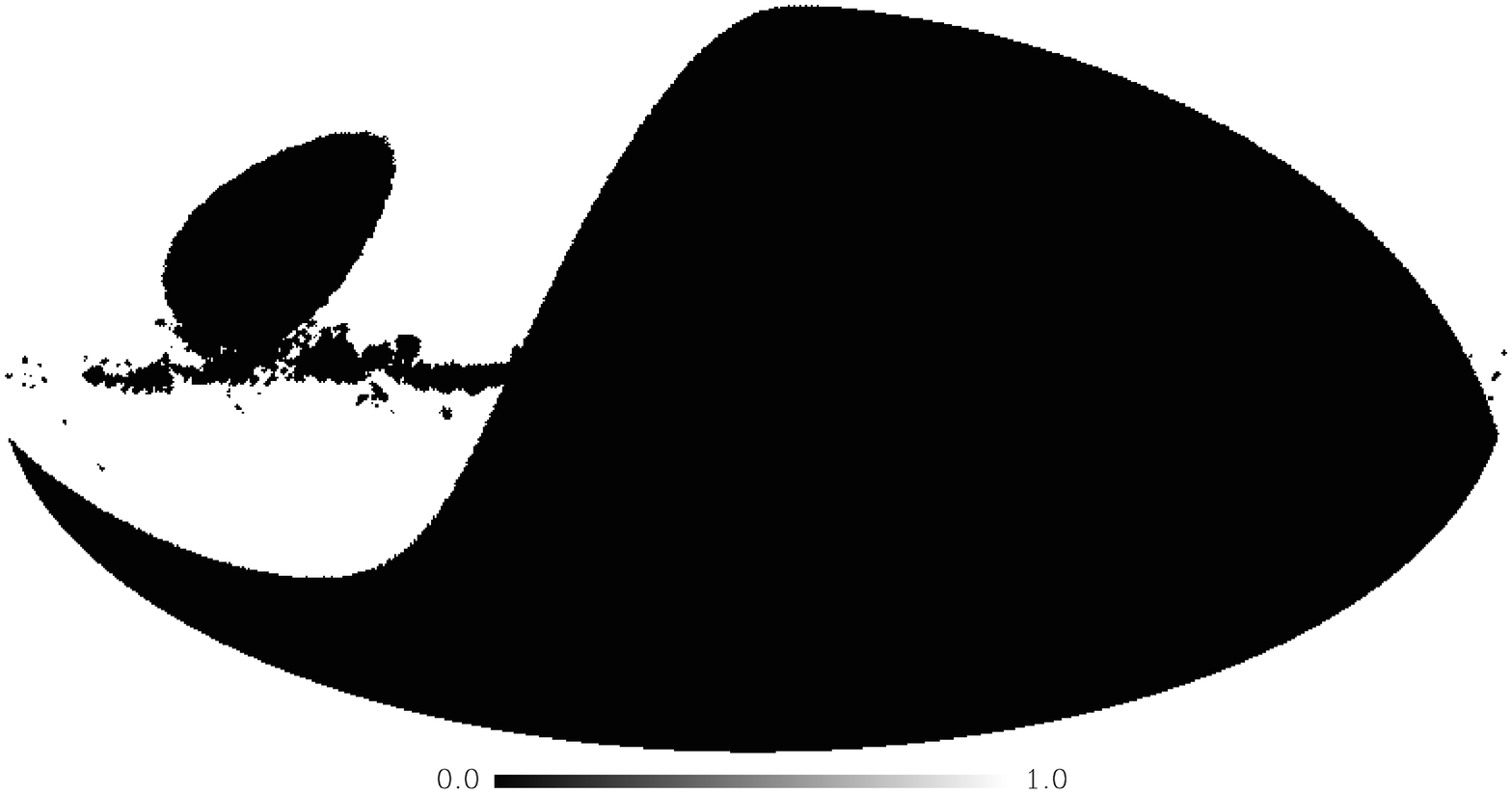}
\caption{Sky masks used for {\it Planck} and COrE (left) and LSPE
  (right). The {\it Planck} and COrE mask is the P06 mask prepared by
  WMAP; it covers around 70\,\% of the sky. The mask for LSPE takes
  into account the sky coverage of the balloon and features a smaller
  Galactic mask, resulting in a final sky coverage of 25\,\%.}
\label{fig:masks}
\end{center}
\end{figure*}
\subsection{Sky model}
We consider three polarised components: the CMB and diffuse polarised Galactic synchrotron and dust emission. The polarized CMB simulation is based on a standard  $\Lambda$CDM model with best-fit cosmological parameters from {\it Planck} \citep[including WMAP polarization,][]{2013arXiv1303.5076P}. We have added tensor modes with tensor to scalar ratio $r=0.2$ and $r=0.1$ and gravitational lensing. The power spectra have been computed with CAMB \footnote{\tt http://camb.info/}.  

The polarization Q and U synchrotron and dust templates were generated
at 100 GHz using the Planck Sky Model
\citep{2013A&A...553A..96D}. These templates were then extrapolated to
lower and higher frequencies using the spectra of
eqs.~(\ref{scaling_synchro}) and (\ref{scaling_dust}) with $\beta_{\rm
  s}=3$ for synchrotron and $\beta_{\rm d}=1.7, \, T_{\rm d}=18\,$K
for dust. These equations and parameters define the true mixing matrix
of the model, which appears in eq.~(\ref{deltacomp}) for the
computation of the foreground residuals.  As mentioned earlier, the
spectral properties have been taken to be spatially constant in order to be able to perform the forecast at the power spectrum level.

\begin{figure*}
\begin{center}
\includegraphics[width=8.8cm,keepaspectratio]{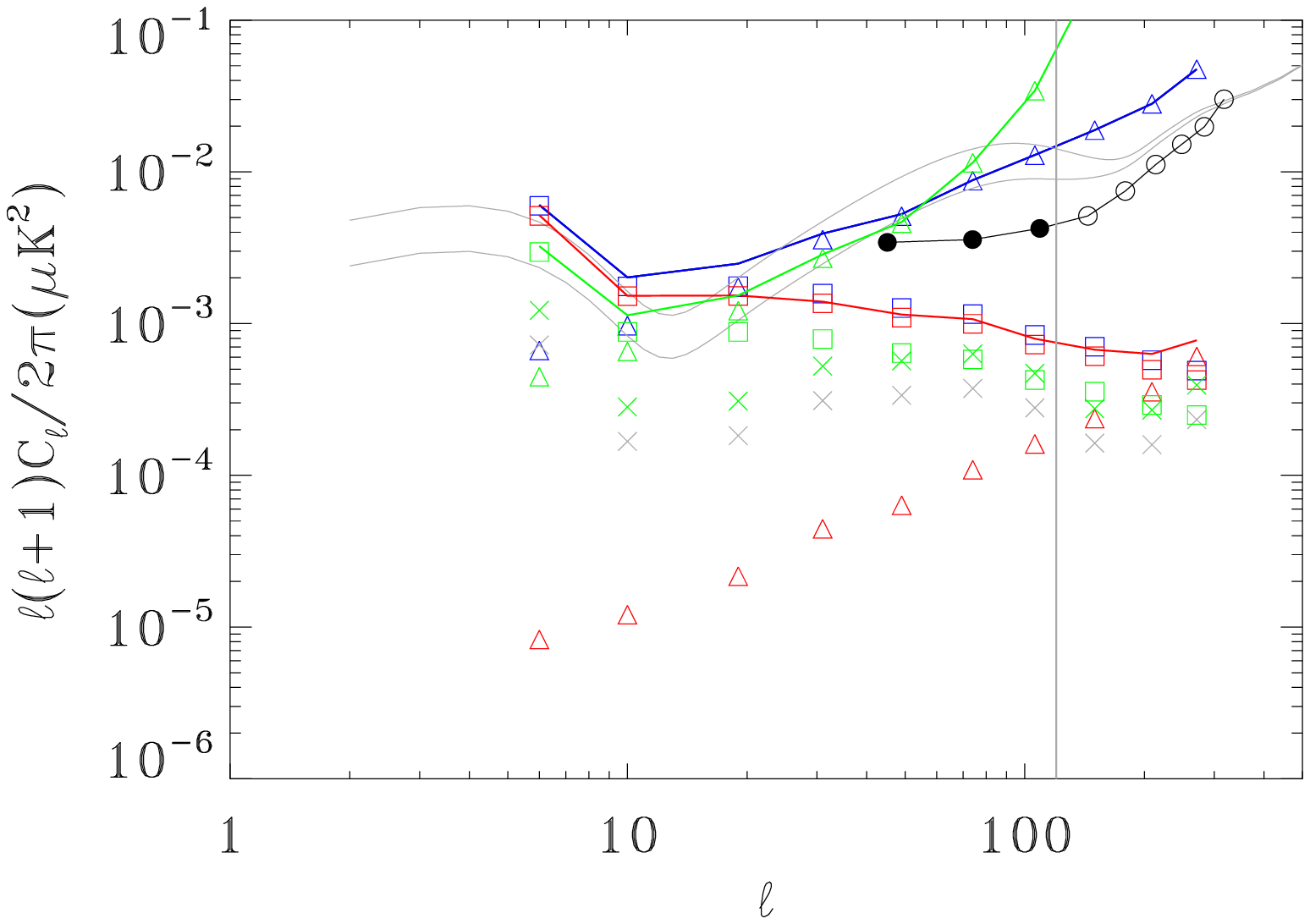}
\includegraphics[width=8.8cm,keepaspectratio]{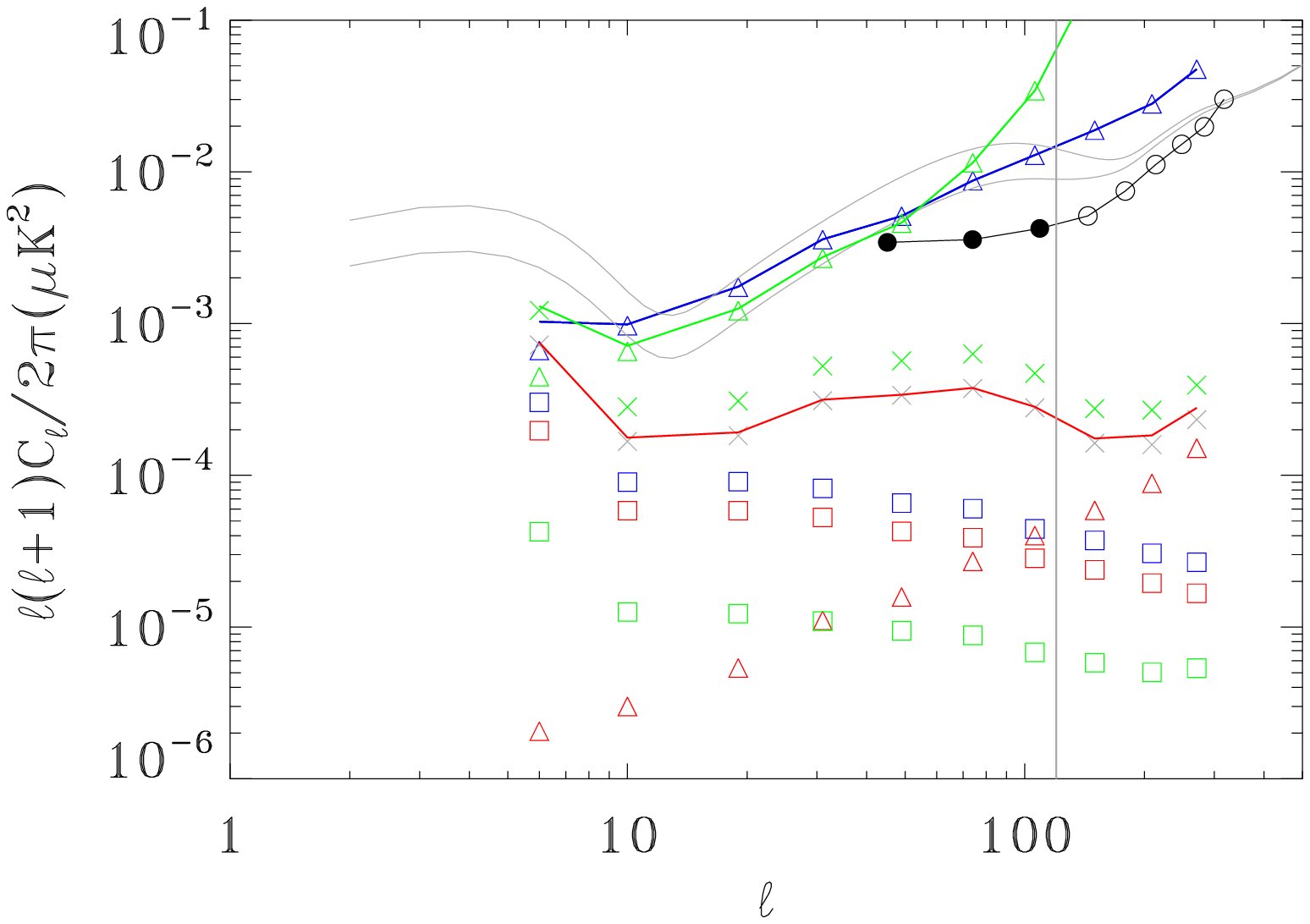}
\caption{Forecasted errors on the $B$-mode power spectrum for
  conservative (left) and improved (right) errors on the mixing
  matrix. Triangles: $\Delta \bmath{C}^{\rm BB}_{\hat \ell,{\rm noise}}$; squares: $\Delta \bmath{C}^{\rm BB}_{\hat \ell,{\rm foreg}}$; crosses: $\Delta \bmath{C}^{\rm BB}_{\hat \ell,{\rm CV}}$ for $r=0.2$ for the mask used
  for LSPE (green) and the one used for {\it Planck} and COrE
  (grey). Solid lines: total error ($\Delta \bmath{C}^{\rm BB}_{\hat \ell}=\Delta \bmath{C}^{\rm BB}_{\hat \ell,{\rm noise}}+\Delta \bmath{C}^{\rm BB}_{\hat \ell,{\rm foreg}} + \Delta \bmath{C}^{\rm BB}_{\hat \ell,{\rm CV}}$ for $r=0.2$). Blue: {\it Planck}; green: LSPE; red:
  COrE. Solid grey lines: theoretical $BB$ power spectra (primordial +
  lensing) for $r=0.1$ and $r=0.2$. The black points show the total
  error (statistical plus cosmic variance) for BICEP2; the filled
  points are those we used in our analysis. The grey vertical line
  indicates the maximum $\ell$ that we included in the Fisher matrix
  analysis for $r$ and $\tau$. This maximum $\ell$ was chosen in order
  to avoid the dominant lensing $B$-mode signal at higher multipoles.}
\label{fig:deltacl}
\end{center}
\end{figure*}
\subsection{Masks}
For the full-sky {\it Planck} and COrE experiments we adopted the WMAP P06 mask \citep{2007ApJS..170..335P}, covering roughly 70\,\% of the sky. The LSPE is a stratospheric baloon experiment and its coverage is limited. In this case we apply a smaller foreground mask to restrain cosmic variance, which is the dominant source of error at large scales. The final coverage is 25\,\% of the sky. The two masks are shown in Fig.~1.

\subsection{Foreground residuals}\label{sec:deltabeta}

As mentioned earlier, the foreground residuals depend on the mismatch between the true and the estimated mixing matrix, which is parametrised by the error on the synchrotron and dust spectral indices, $\Delta \beta_{\rm s }$ and $\Delta \beta_{\rm d}$. 

For the mixing matrix error we considered two regimes, which we label
as ``conservative'' ($\Delta \beta_{\rm s}=0.1$ , $\Delta \beta_{\rm
  d}=0.05$) and ``improved'' ($\Delta \beta_{\rm s}=0.01$ , $\Delta
\beta_{\rm d}=0.005$). The former are a conservative assessment of the
state-of-the-art, based on a realistic simulation of {\it Planck}
polarization data, including a spatially-varying mixing matrix and
realistic noise (Ricciardi et al. 2010). The next generation
experiments which are focused on detailed characterisation the $B$-mode power
spectrum will undoubtedly improve the accuracy of the determination of
foreground spectral properties. In order to forecast the performance
of these experiments we therefore also consider an ``improved'' regime. The adopted error values should be considered as indicative rather than representative. The error on the dust spectral index is smaller than that on the synchrotron spectral index because of the frequency coverage of the experiments, which is broader towards high frequency, thus providing in principle better control over dust contamination. 

To propagate the errors $\Delta \beta_{\rm s }$ and $\Delta \beta_{\rm d}$ to the CMB power spectra, we generated a set of ten estimated mixing matrices $\bmath{\sf {\hat H}}$, drawing the actual spectral indices from Gaussian distributions with mean equal to the true indices and standard deviation equal to the spectral index errors. For each  $\bmath{\sf {\hat H}}$ we computed the reconstruction matrix $\bmath{\sf W}$, the residual map through eq.~(\ref{deltacomp}) and its power spectrum outside the adopted mask. 

For the CMB reconstruction we considered the frequency channels in the
range $40<\nu<250$\,GHz. As discussed in Bonaldi \& Ricciardi (2011),
using a wider frequency range could potentially result in increased
foreground residuals due to the inclusion of frequency channels which
are more affected by foregrounds. On the other hand, including further
channels generally lowers the noise: the best trade-off for a
particular experiment will depend on the instrument sensitivity, on
the intensity of the foregrounds, and on the errors on the mixing
matrix. We stress that, for the purposes of estimating the mixing
matrix, it is always beneficial to use as wide a frequency range as
possible. In this respect, the lowest and highest frequencies are
particularly useful, as they map respectively the polarized synchrotron and dust emission with high signal-to-noise and low contamination from the other components. Indeed, the accuracy of the mixing matrix estimation is critically important, particularly for the high sensitivity instruments.

\subsection{Instrumental specifications}\label{sec:inst_spec}
In this work we consider the {\it Planck} and COrE instruments, as representative of a current and future CMB satellite. We also consider LSPE, as representative of a balloon-borne experiment targeting the large-scale polarization of the CMB. 
For the purpose of our forecasts, the instruments are modelled as a set of frequency channels, each one having a given
resolution and sensitivity (which we take to be uniform across the sky). As
previously discussed, we assume that the CMB is separated from the
foreground contamination by way of a linear combination of the frequency maps between 40 and 250\,GHz. Therefore, we do not explicitly consider frequency channels beyond this range. The instrumental specifications adopted are reported in Table~\ref{tab:instr}, together with the references we used to derive them.  

We have also included the BICEP2 $EE$ and $BB$ power spectra constraints
in our forecasts in particular to assess the potential improvement
achievable by combining them with another measurement at larger angular
scales. We downloaded the BICEP2 data and considered the statistical
error bars for the $EE$ and $BB$ power spectra. We added to them the
uncertainties due to cosmic variance for a sky coverage of 380 square
degrees. For $BB$ we used only the first three band powers (covering
$\ell=45$--110). In fact, these are the most important bands for
constraining the tensor-to-scalar ratio $r$, because at higher $\ell$s
the contribution due to lensing $B$-modes is dominant. 

\section{Results}\label{sec:results}
In Figure \ref{fig:deltacl} we show a comparison of the theoretical power spectra for $r=0.1$ and 0.2 with the forecasted error bars for each of the considered experiments and both conservative (left) and improved (right) mixing matrix errors. 

The large-scale $B$-mode signal for these values of the tensor-to-scalar ratio is accessible to both {\it Planck} (blue lines and symbols) and LSPE (green lines and symbols). For {\it Planck} the limiting factor is foreground residuals at very low multipoles and noise at higher multipoles. LSPE is limited by foreground residuals and cosmic variance at very low multipoles and noise at higher multipoles. In both cases, the different mixing matrix accuracy from the conservative to the improved case lowers the total error only for a limited multipole range. 
Since the total error for {\it Planck} and LSPE are quite similar for
scales larger than those probed by BICEP2, the combination of either
of these probes with the BICEP2 points results in a similar constraint on the cosmological parameters. Therefore, in the Fisher matrix analysis that follows, we only show the results for {\it Planck}.

The performances of the COrE experiment (red lines and symbols) are
limited by the accuracy of the mixing matrix estimation for a large
multipole range. Thus the results for COrE illustrate the impact of
the improvement in the mixing matrix accuracy when going from the conservative errors to the improved ones. This experiment is able to measure accurately both the reionization bump and the main peak. 

\begin{figure*}
\begin{center}
\includegraphics[width=8.5cm,keepaspectratio]{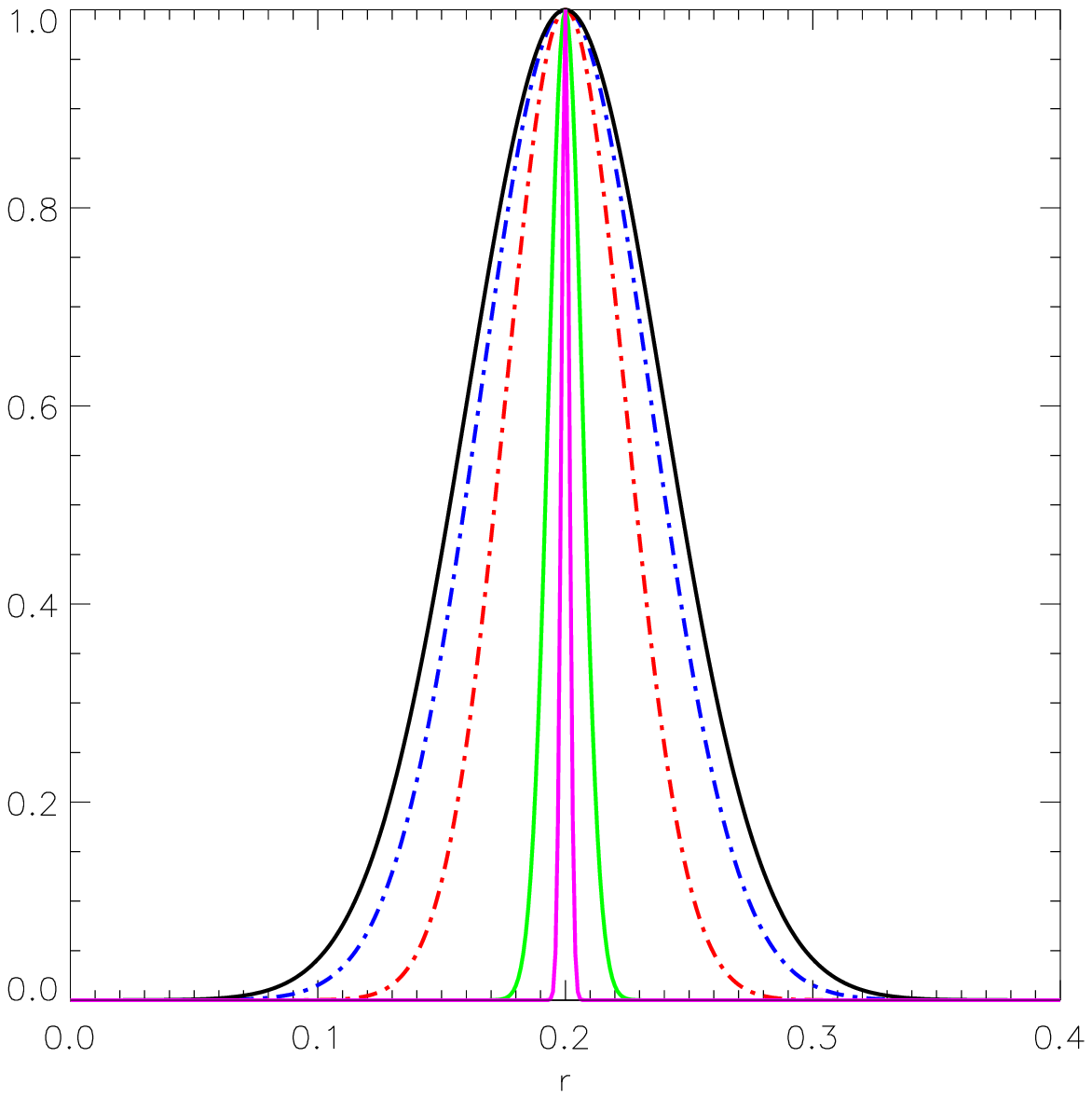}
\includegraphics[width=8.5cm,keepaspectratio]{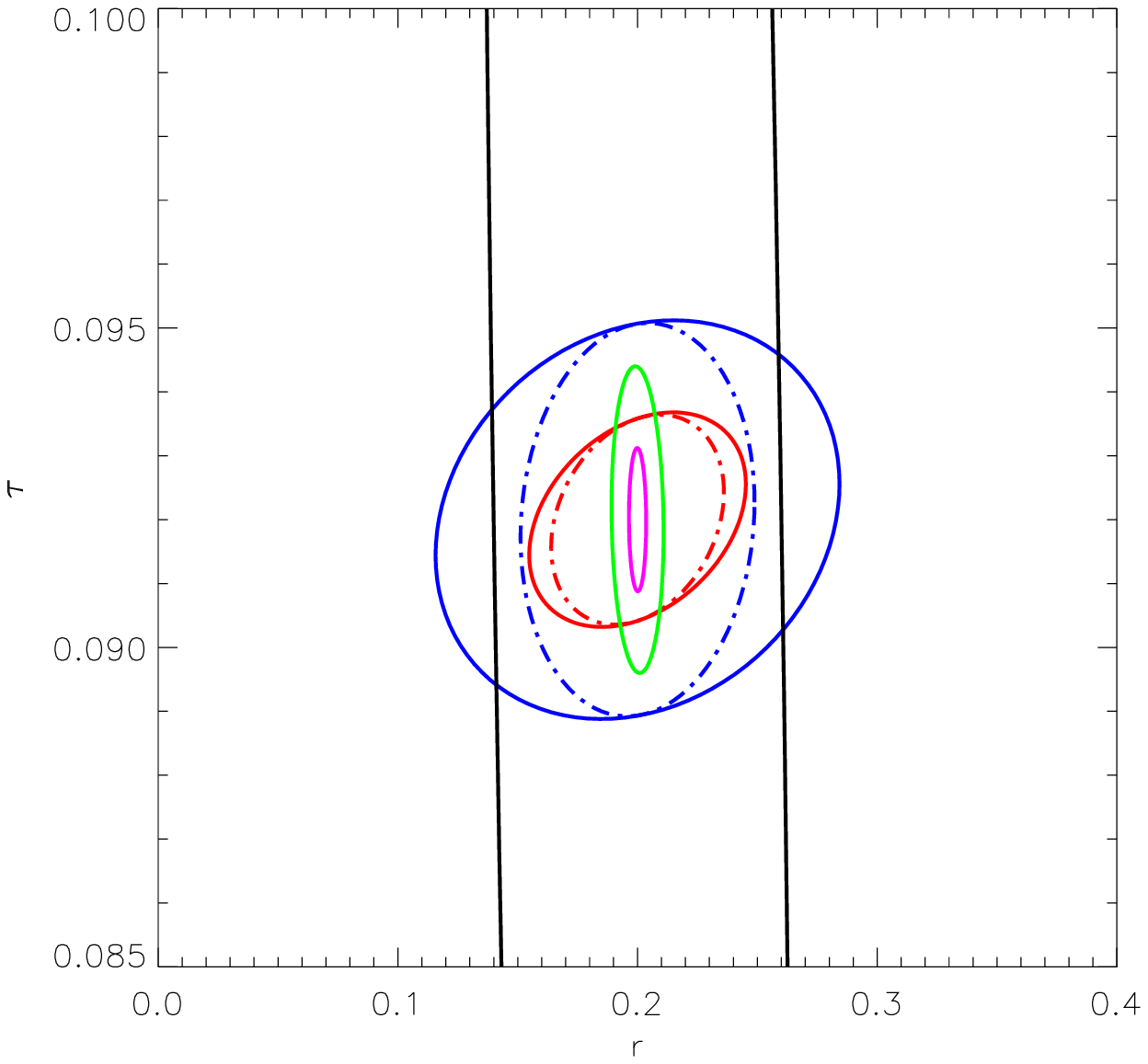}
\caption{One-dimensional likelihood for $r$ (left) and two-dimensional likelihood for $r$ and $\tau$ (right) when $r=0.2$ and all the other parameters are known. The lines and contours represent the 68\% CL (1\,$\sigma$). Solid lines represent the experiments used alone, and dot-dashed lines in combination with the BICEP2 constraint. Black: BICEP2; blue: {\it Planck} with conservative mixing matrix errors; red: {\it Planck} with improved mixing matrix errors; green: COrE with conservative mixing matrix errors; magenta: COrE  with improved mixing matrix errors. The results for LSPE are similar to those obtained for {\it Planck}.}
\label{fig:fisher}
\end{center}
\end{figure*}

In the left panel of Fig.~\ref{fig:fisher} we show the results of the
Fisher matrix analysis when the only free parameter is $r$. This
corresponds to assuming that all the other parameters are known. For
this analysis we used only the $BB$ spectrum up to $\ell=100$. In
order to exploit measurements at higher multipoles, one would also
need to consider the amplitude of lensing parameter, $A_{\rm L}$.

If $r=0.2$, our Fisher matrix analysis of the BICEP2 results predicts
$\Delta r=0.04$ at 68\% confidence level (CL). The somewhat lower
error with respect to the published result ($r=0.2^{+0.07}_{-0.05}$)
is partly intrinsic to the Fisher matrix approach (Cramer-Rao
inequality) and partly due to our simplified treatment of the BICEP2
errors, such as having neglected the correlation of errors between
different bins. In the following we will use the predicted BICEP2
errors in place of the actual errors in order to compare the different
probes using the same analysis. 

In the simple one-parameter model, the inclusion of the {\it Planck}
dataset at low multipoles improves the constraint on $r$ by 25--45\%
depending on the mixing matrix accuracy. The results for LSPE are very
similar. Besides this improvement on the accuracy, these experiments
would be able to provide an independent confirmation of the primordial
origin of the BICEP2 signal were it to be cosmological. A next generation experiment like COrE would then improve the measurement of $r$ substantially (almost one order of magnitude), especially if the mixing matrix is accurately modelled.  

In the right panel of Fig.~\ref{fig:fisher} we consider the
simultaneous estimation of $r$ and $\tau$ using both the $EE$ and $BB$
power spectra. BICEP2 alone (black line) measures $r$ much better than
$\tau$, because it does not measure the reionization signal at large angular scales.

{\it Planck} alone (blue and red solid lines for the conservative and improved mixing matrix errors) has a better handle on the combination of these parameters, but the results on $r$ are somewhat worse than the BICEP2 ones. The combination of the two probes (blue and red dot-dashed lines) gives the best results and measures both $r$ and $\tau$ accurately. With respect to BICEP2 alone, the improvement on $\tau$ is $\sim$95--100\%. 

In the conservative error regime, COrE reduces the error bar on $r$ by
80\% and on $\tau$ by 95\% with respect to BICEP2 (green line). If the
error bars on the mixing matrix are those of the improved regime, the
error bars on both parameters are reduced by almost two orders of
magnitudes (magenta line). This demonstrates the importance of a detailed understanding of the foreground spectra for the next generation experiments. 

\section{Conclusions}
We have applied the method developed by Bonaldi \& Ricciardi (2011) to
forecast error bars on the CMB polarization $EE$ and $BB$ power
spectra for current and future experiments, in the hypothesis that
$r=0.1$--0.2 as suggested by the BICEP2 experiment. We showed that
such a signal is within {\it Planck}'s reach even for conservative assumptions on the accuracy of foreground removal (but without considering instrumental systematics). The detection of the large-scale counterpart of the BICEP2 signal would be the most convincing confirmation of this result. 

We used a Fisher matrix formalism to predict the errors on the
cosmological parameters starting from the error bars on the $BB$ power
spectrum. The combination of  BICEP2 with either {\it Planck} or with a balloon-borne experiment targeting the large-scale polarization of the CMB (here represented by LSPE) improves the accuracy on $r$ by 25-45\% and measures $\tau$ with an error of 0.002--0.001. 

The constraint on the tensor-to-scalar ratio can be improved
substantially with a next-generation $B$-mode satellite such as
COrE. This experiment can reduce the error bar on $r$ by another order
of magnitude, provided that we have accurate knowledge of the
frequency spectra of the foreground components. On the other hand, if
only limited progress on this aspect is made from the present state-of-the-art, the improvement in sensitivity with respect to {\it Planck} cannot be fully exploited. This confirms that, even for a relatively large $B$-mode signal such as the one implied by the BICEP2 results, foreground removal is crucial for a precise measurement of the tensor-to-scalar ratio. 

\section{Aknowledgements}
We thank the referee, F. Stivoli, for useful suggestions.
AB and MLB acknowledge support from the European Research Council under the EC FP7 grant number 280127. MLB also acknowledges support from an STFC Advanced/Halliday fellowship. SR acknowledges support by ASI through ASI/INAF Agreement 2014-024-R.0
for the Planck LFI Activity of Phase E2 and by MIUR through PRIN 2009 grant no. 2009XZ54H2.
\bibliographystyle{mn2e}
\bibliography{biblio}

\end{document}